\documentclass[12pt]{article}
\usepackage{smile}
\usepackage{amsmath}
\usepackage{amsthm}
\theoremstyle{remark}
\usepackage{graphicx}
\usepackage{enumerate}
\usepackage{natbib}
\usepackage{url} 
\usepackage{algorithm}
\usepackage{algorithmic}
\usepackage{natbib}
\usepackage{comment}
\usepackage{setspace}
\usepackage{graphics,latexsym,amssymb,epsfig,bm,setspace,amsmath,textcomp,url,xcolor,graphicx}
\usepackage[colorlinks, allcolors=black, linkcolor=black,anchorcolor=black,filecolor=black,citecolor=black]{hyperref}

\usepackage{bbm}

\newcommand\blfootnote[1]{%
  \begingroup
  \renewcommand\thefootnote{}\footnote{#1}%
  \addtocounter{footnote}{-1}%
  \endgroup
}

\newcommand\my[1]{{\color{cyan}{[MY: #1]}}}

\begin{document}
\title{Generalized Tensor Completion with Non-Random Missingness}
\date{}
\author{%
{Maoyu Zhang$^{1^*}$, Biao Cai$^{2^*}$, Will Wei Sun$^3$ and Jingfei Zhang$^1$}
\vspace{1.6mm}\\
\fontsize{11}{10}\selectfont\itshape $^1$\,Goizueta Business School, Emory University, Atlanta, GA, USA. \\
\fontsize{11}{10}\selectfont\itshape $^2$\,Department of Decision Analytics and Operations, City University of Hong Kong, Hong Kong. \\
\fontsize{11}{10}\selectfont\itshape $^3$\,Daniels School of Business, Purdue University, West Lafayette, IN, USA. \\
\blfootnote{The first two authors contributed equally to this work.}}

\renewcommand{\baselinestretch}{1.15}
\maketitle
\begin{abstract}
Tensor completion plays a crucial role in applications such as recommender systems and medical imaging, where data are often highly incomplete. While extensive prior work has addressed tensor completion with data missingness, most assume that each entry of the tensor is available independently with probability $p$. However, real-world tensor data often exhibit missing-not-at-random (MNAR) patterns, where the probability of missingness depends on the underlying tensor values. This paper introduces a generalized tensor completion framework for noisy data with MNAR, where the observation probability is modeled as a function of underlying tensor values. Our flexible framework accommodates various tensor data types, such as continuous, binary and count data. For model estimation, we develop an alternating maximization algorithm and derive non-asymptotic error bounds for the estimator at each iteration, under considerably relaxed conditions on the observation probabilities. Additionally, we propose a statistical inference procedure to test whether observation probabilities depend on underlying tensor values, offering a formal assessment of the missingness assumption within our modeling framework. The utility and efficacy of our approach are demonstrated through comparative simulation studies and analyses of two real-world datasets.
\end{abstract}


\newpage
\baselineskip=26.5pt
\section{Introduction}\label{sec:intro}

Tensor completion is an important problem in modern data analysis, with broad applications in areas such as recommender systems, medical imaging, and social network analysis. The objective of tensor completion is to recover the entire tensor from noisy and partially observed entries, typically under a low-rank assumption \citep{kolda2009tensor}. For example, in recommender systems, user-item-context interactions can be represented as a third-order tensor. In practice, only a small fraction of these interactions are observed, and tensor completion techniques are used to predict missing user preferences across different contexts.

\begin{figure}[!ht]
\centering
\includegraphics[width=\linewidth]{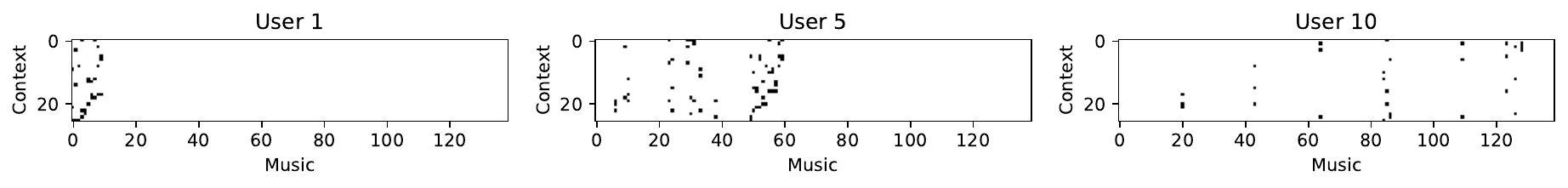}
\caption{Missingness patterns for three users from the InCarMusic dataset. Marked cells indicate observed entries, while blank cells indicate missing values.}
\label{fig:music}
\end{figure}

Over the past decade, a wide range of methods have been proposed to address the tensor completion problem and provide theoretical guarantee, including \cite{kolda2009tensor,liu2012tensor,krishnamurthy2013low,yuan2016tensor,bi2018multilayer,lee2020tensor,xia2021statistically,cai2019nonconvex,ibriga2023covariate,ma2024statistical}, among many others. 
The majority of existing approaches assume that the entries are missing uniformly at random, a special case of the missing completely at random (MCAR) assumption. While this assumption simplifies both modeling and theoretical analysis, it is often unrealistic in practice. 
Figure \ref{fig:music} shows the missingness mask (i.e., binary indicators of whether each entry is observed) for three users in the InCarMusic dataset, which contains ratings for 139 songs by 42 users across 26 contextual conditions (see Section \ref{sec:real} for detail). It is seen that users typically only rate a subset of songs across the various contexts.

In recommender systems, it is common for users to only rate items they strongly like or dislike, making the probability of observing a rating dependent on its underlying preference value, a setting known as missing not at random (MNAR). 
When missingness depends on the underlying values, standard tensor completion methods that ignore this mechanism can yield biased estimates of the true tensor. For example, if users are more likely to rate items they like, observed data will over-represent high ratings, causing the model to over-estimate the underlying user preference. In such cases, failing to properly account for the MNAR mechanism in tensor completion can lead to biased estimates and invalid statistical inference.

In this paper, we propose a generalized tensor completion framework that allows for data missing not at random. We model the observation probabilities as a function of underlying tensor values and adopt a joint modeling approach for the observed entries and the missingness mask. 
Our framework is flexible and accommodates a variety of tensor data types, including continuous, binary, and count data, and also includes tensor completion with uniform missing as a special case. 
For estimation, we develop an alternating maximization algorithm that is easy and fast to implement.
We further develop theoretical guarantees for the proposed method. 
First, we derive explicit error bounds for the estimates from each step of the proposed algorithm. This result reveals an interesting interplay between computational and statistical errors. As the number of iterations increases, the computational error of the estimates decreases geometrically until it reaches a neighborhood within the statistical error of the true parameter. Our results are established under much weaker conditions on the observation probability tensor, requiring only conditions on slice-wise averages, whereas existing results typically assume uniform bounds on all probabilities entries.
A main challenge in our analysis, compared to prior work, is that we allow observation probabilities to depend on underlying tensor values and to be highly heterogeneous. Probabilities may approach 0 or 1 at different rates, making it challenging to derive tight bounds for the curvature of the objective function. In particular, when observation probabilities are allowed to be arbitrarily close to 0 or 1, the uniform boundedness assumption commonly imposed in prior work does not hold, and direct strong concavity arguments lead to inadequate error bounds. To address this, we consider a new proof strategy based on analyzing individual coordinates of the low-rank components instead of the usual vector-based analysis, which provides a sharper control of the statistical error.

Besides heterogeneous probabilities, our analysis is further complicated by the nonconcavity of the objective function and nonlinearities in models for both the tensor entries and the observation probabilities. Unlike standard tensor completion with squared loss, both the tensor data model and the missingness mechanism in our framework are generalized linear models, so parameter updates have no closed form and require delicate control of gradients and Hessians.
Under our modeling framework, we also introduce a statistical testing procedure to assess whether the observation probability depends on the underlying tensor values, offering a formal assessment of the data missingness mechanism. One useful observation is that tensor completion and observation probability modeling have different sample complexity requirements, with the latter requiring much fewer samples. Based on this insight, we design a sample splitting approach that ensures errors from estimating the low-rank tensor is negligible when deriving the asymptotic distribution of the parameters in modeling observation probabilities.

\subsection{Related literature}


The first line of related research investigates low-rank tensor decomposition \citep{kolda2009tensor,anandkumar2014tensor,sun2017provable,zhang2018tensor,wang2020learning, chen2025distributed, xu2025statistical}, which aims to recover latent low-rank structures from observed tensor data. Building on these foundations, a subsequent body of work studies low-rank tensor completion under uniform missingness, where each tensor entry is observed independently with equal probability \citep{kolda2009tensor,liu2012tensor,yuan2016tensor,xia2021statistically,cai2019nonconvex,ibriga2023covariate,ma2024statistical}. While these methods are effective for exploiting the tensor low-rank structure, they are designed for uniformly missing data and do not address the more challenging setting of MNAR, which is the focus of this paper.

The second line of related research considers matrix and tensor completion with non-uniform missing. That is, entries are observed with unequal probabilities, but these probabilities are assumed to be independent of the underlying tensor values. 
\cite{mao2019matrix, mao2021matrix, mao2024mixed} develop covariate-assisted and inverse propensity scoring-based frameworks assuming data missing at random (MAR). \cite{chao2021hosvd} propose a weighted tensor HOSVD algorithm that accounts for heterogeneous sampling. \cite{lee2020tensor} study ordinal tensor completion under non-uniform sampling, and \cite{mo2025act} introduce a clustering-based tensor completion framework for financial panels with severe heterogeneous missingness. These works allow heterogeneous missingness but assume independence from the underlying data values.

The third and most closely related line of research focuses on matrix and tensor completion under MNAR mechanisms. \cite{ma2019missing,bhattacharya2022matrix} consider a two-step procedure that first estimates observation probabilities and then performs matrix completion.
Both works require the observation probabilities to be bounded away from both 0 and 1 by a constant. 
\cite{athey2021matrix,bai2021matrix,chernozhukov2023inference,choi2024matrix} consider matrix completion with structured missing such as block missing in causal data panels.
\cite{agarwal2023causal} propose a causal framework that allows missingness to depend on both the underlying values and other missing entries. However, they impose 
a strong assumption on the number and location of observed entries. 
For tensor completion under MNAR mechanisms, \cite{yang2021tenips} consider a similar framework as in \cite{ma2019missing}. In addition, they consider a noiseless setting and assume the tensor of observation probabilities is low-rank, which can be restrictive. 
More importantly, the inverse propensity score (IPS) estimation employed in \cite{ma2019missing,yang2021tenips} can produce large estimation errors when propensity scores are very close to zero.

\subsection{Notation}\label{sec:notation}
We denote vectors by lower-case bold letters (e.g., $\bx$), matrices by upper-case bold letters (e.g., $\bX$), and tensors by upper-case script letters (e.g., $\cX$). For $d\in\mathbb{N}^+$, let $[d]=\{1,2,\dots,d\}$.
Given a vector $\bx\in\mathbb{R}^d$, we use $\Vert\bx\Vert_2$ to denote its $\ell_2$ norm. 
Let $\langle{\ba},{\bb}\rangle=\sum_i a_ib_i$ for ${\ba},{\bb}\in\mathbb{R}^d$, $\circ$ denote the vector outer product and $*$ the entrywise product. 
For a tensor $\cX \in \mathbb{R}^{d_1\times d_2\times d_3}$, {let $\cX_{ijk}$ denote its $(i,j,k)$th entry, $\cX_{ij\cdot}$ the $(i,j)$th tube fiber, and $\cX_{\cdot\cdot k}$ the $k$th frontal slice; let $\|\cX\|_F$ denote its Frobenius norm.
We write $a=\Theta(b)$ if there exist constants $0 < c_1 \le c_2 < \infty$ such that $c_1 |b| \le |a| \le c_2 |b|$.

The remainder of the paper is organized as follows. Section \ref{sec:model} introduces the generalized tensor completion framework with MNAR, and describes the model estimation procedure. Section \ref{sec:theo} presents theoretical properties of the proposed estimator including a hypothesis testing procedure. Section \ref{sec:sim} reports simulation results, and Section \ref{sec:real} conducts analyses of two real-world data sets. The paper is concluded with a short discussion section.

\section{Model and Estimation}\label{sec:model}
Let $\cX\in \mathbb{R}^{d_1\times d_2\times d_3}$ denote the unknown tensor of interest, which is not observed directly. Instead, we observe entries of a data tensor $\cY \in\mathbb{R}^{d_1\times d_2\times d_3}$. Given $\cX$, the entries $\cY_{ijk}$'s are independent, and each $\cY_{ijk}$ follows an exponential family distribution with density
\begin{equation}
\label{eqn:exponential_family}
f(\cY_{ijk} \mid \cX_{ijk}) = c(\cY_{ijk})\exp\Big(\frac{\cY_{ijk} \cX_{ijk} - \psi(\cX_{ijk})}{\phi_0}\Big),
\end{equation}
where $c(\cdot)$ is the normalizing factor, $\psi(\cdot)$ is the cumulant function and $\phi_0>0$ is a known parameter. 
Under \eqref{eqn:exponential_family}, the conditional mean is given by $\mathbb{E}[\cY_{ijk}|\cX_{ijk}]=h^{-1}(\cX_{ijk})$, where $h(\cdot)$ is an invertible link function such that $h^{-1}(\cdot)=\psi'(\cdot)$.

In real applications, not all entries of $\cY$ can be observed. Define a binary mask tensor $\cD \in \{0,1\}^{d_1\times d_2\times d_3}$, where $\cD_{ijk}=1$ if $\cY_{ijk}$ is observed and $0$ otherwise. Given $\cX$, we assume the entries $\cD_{ijk}$'s are independent and
$$
\cD_{ijk}\mid \cX_{ijk} \sim \mathrm{Bernoulli}(\cP_{ijk}), 
\qquad \cP_{ijk}=g_{\btheta}(\cX_{ijk}),
$$
where $g_{\btheta}:\mathbb{R}\rightarrow [0,1]$ is a known smooth function with unknown parameter $\btheta$. When $g_{\btheta}(\cdot)$ is constant, the missing mechanism reduces to MCAR. Allowing $g_{\btheta}(\cdot)$ to vary with $\cX_{ijk}$ gives the more flexible MNAR mechanism. A common specification \citep{ma2019missing,yang2021tenips} considers
\begin{equation}
g_{\btheta}(\cX_{ijk}) = \text{logit}^{-1} (b_0 + b_1 \cX_{ijk}),
\end{equation}
where $b_0$ is a baseline parameter and $b_1$ characterizes how the unobserved entry $\cX_{ijk}$ alters its probability of being observed. More generally, $g_{\btheta}(\cX_{ijk})$ can include polynomials of $\cX_{ijk}$ and possibly additional covariates, to flexibly model missingness mechanisms.

Let $\Omega = \{(i,j,k): \cD_{ijk}=1\}$ denote the index set of observed entries in $\cY$, 
and write $\cY_{\mathrm{obs}}= \{\cY_{ijk} : (i,j,k)\in\Omega\}$.
The joint density of $\cY_{\mathrm{obs}}$ and $\cD$ follows
\begin{align}
f(\cY_{\text{obs}}, \cD \mid \cX) 
&= f(\cY_{\text{obs}} \mid\cD,\cX) \, f(\cD \mid \cX) \label{eqn:joint_likelihood} \\
&= \prod_{i,j,k \in \Omega} f(\cY_{ijk} \mid \cD_{ijk},\cX_{ijk}) \, \prod_{i,j,k}\cP_{ijk}^{\cD_{ijk}} (1-\cP_{ijk})^{1-\cD_{ijk}}.\nonumber
\end{align}

Given observed tensor $\cY_{\text{obs}}$ and missingness mask $\cD$, our goal is to estimate the underlying tensor $\cX$ via \eqref{eqn:joint_likelihood}. 
We assume that the underlying true tensor $\cX$ admits a rank-$R$ CP decomposition \citep{kolda2009tensor}:
\begin{equation}
\cX = \sum_{r=1}^{R} \lambda_r \bu_{r}\circ\bv_{r}\circ \bw_{r},
\label{eqn:CP}
\end{equation}
where $\lambda_r\in\mathbb{R}^+$, $\bu_{r}\in\mathbb{R}^{d_1}$, $\bv_{r}\in\mathbb{R}^{d_2}$ and {$\bw_{r}\in\mathbb{R}^{d_3}$}.
For identifiability, $\bu_{r}$'s, $\bv_{r}$'s and $\bw_{r}$'s are assumed to be unit length vectors, that is, $\|\bu_{r}\|_2=\|\bv_{r}\|_2=\|\bw_r\|_2=1$ for any $r\in[R]$. 
This CP low-rank structure 
is commonly considered in tensor problems, such as tensor completion \citep{wang2020learning,bi2018multilayer,lee2021beyond,tao2024efficient}, tensor regression \citep{sun2017provable,zhou2023partially}, tensor bandits \citep{zhou2025stochastic}, tensor clustering \citep{cai2024jointly}, and covariance decomposition \citep{deng2023correlation}.

Next, we compare our joint estimation approach with the commonly employed two-step matrix and tensor completion methods for MNAR settings \citep{ma2019missing,yang2021tenips,bhattacharya2022matrix}. 
The two-step approaches first estimate observation probabilities from the missingness mask, typically under a low-rank assumption, and then perform matrix or tensor completion after IPS reweighting.
When propensity scores (i.e., observation probabilities) are close to zero, these methods often suffer from large estimation errors. Indeed, their theoretical analyses often require the observation probabilities to be uniformly bounded away from 0 and 1. 
In contrast, our method jointly estimate $\cX$ and $\btheta$ by leveraging both the observed data and the missingness mask, without relying on reweighting, and is therefore less sensitive to extreme probabilities.
Our theoretical analysis allows observation probabilities to be arbitrarily close to 0 or 1, provided that the slice-wice averages are not too extreme (see Condition 3 in Section \ref{sec:theo}). 
Furthermore, our framework enables hypothesis testing of MCAR versus MNAR, which is not available in existing two-step methods.

Next, we discuss model estimation. We collect all unknown parameters as
$$
\bTheta 
= \Big( \{\bu_r^\top\}_{r=1}^R,\; \{\bv_r^\top\}_{r=1}^R,\; \{\bw_r^\top\}_{r=1}^R,\; \{\lambda_r\}_{r=1}^R, \; \btheta^\top\Big)^\top.
$$
Given observations $\cY_{\text{obs}}$ and missingness mask $\cD$, our goal is to estimate $\bTheta$. 

Based on \eqref{eqn:joint_likelihood} and up to a constant, the joint log-likelihood can be written as $\ell_d(\bTheta)$. For example, if observed tensor entry $\cY_{ijk}$ follows a Bernoulli distribution with $h(\cdot)$ as the logit link and $g_{\btheta}(\cX_{ijk}) = \text{logit}^{-1} (b_0 + b_1 \cX_{ijk})$, we have
\begin{equation*}
\begin{aligned}
\ell_d(\bTheta) &=\sum_{i,j,k \in \Omega}\left\{\cY_{ijk}\cX_{ijk}-\log(1+\exp(\cX_{ijk}))\right\}+\sum_{i,j,k \in \Omega}(b_0 + b_1\cX_{ijk})\\
&\quad\quad-\sum_{i,j,k} \log \Big(1 + \exp\big(b_0 + b_1 \cX_{ijk}\big)\Big),
\end{aligned}
\end{equation*}
where $\cX_{ijk}=\sum_{r=1}^{R} \lambda_r u_{ri}v_{rj}w_{rk}$.
We estimate $\bTheta$ by solving the following problem:
\begin{align} \label{eqn:objective}
\max_{\bTheta} \;\; \ell_d(\bTheta), \quad\quad\text{subject to } & \|\bu_r\|_2 = \|\bv_r\|_2 = \|\bw_r\|_2 = 1, \quad r \in [R]. 
\end{align}
We consider an alternating maximization algorithm, as summarized in Algorithm \ref{alg:AGD}.

\begin{algorithm}[!t]
\caption{Alternating maximization algorithm for (\ref{eqn:objective})}
\begin{algorithmic}[1]
\STATE \textbf{input:} data $\cY_{\text{obs}}$, missingness mask $\cD$, rank $R$. 
\STATE \textbf{initialization:}  obtain $\bTheta^{(0)}$ via Algorithm S1. 
\REPEAT
\FOR{$r=1$ to $R$}
\STATE  $\tilde\bu_{r}^{(t+1)} =\arg\max_{\bu_r} \ell_d( \bu_r, \bTheta_{-\bu_r}^{(t)})$, \quad $\bu_{r}^{(t+1)}=\tilde\bu_{r}^{(t+1)}/\|\tilde\bu_{r}^{(t+1)}\|_2$,
\STATE  $\tilde\bv_{r}^{(t+1)}=\arg\max_{\bv_r}   \ell_d(\bv_r, \bTheta_{-\bv_r}^{(t)}$),\quad
$\bv_{r}^{(t+1)}=\tilde\bv_{r}^{(t+1)}/\|\tilde\bv_{r}^{(t+1)}\|_2$,
\STATE $\tilde\bw_{r}^{(t+1)} = \arg\max_{\bw_r} \ell_d(\bw_{r}, \bTheta_{-\bw_r}^{(t)})$,\quad
$\bw_{r}^{(t+1)}=\tilde\bw_{r}^{(t+1)}/\|\tilde\bw_{r}^{(t+1)}\|_2$,
\STATE  $\lambda_{r}^{(t+1)}=\arg\max_{\lambda_r} \ell_d(\lambda_{r}, \bTheta_{-\lambda_r}^{(t)})$, \\
\ENDFOR
\STATE $\btheta^{(t+1)} = \arg\max_{\btheta} \ell_d(\btheta, \bTheta_{-\btheta}^{(t)})$, 
\UNTIL the stopping criterion is met.
\STATE\textbf{output:} $\hat{\bTheta}$.
\end{algorithmic}
\label{alg:AGD}
\end{algorithm}
In Algorithm \ref{alg:AGD}, we use $\bTheta^{(t)}_{-\bu_r}$ to denote $\bTheta$ with $\bu_r$ removed and 
parameters updated before and after $\bu_r$ take values from their $(t+1)$-th and $t$-th steps, respectively. For example, $\bTheta^{(t)}_{-\bu_r}=(\bu_1^{(t+1)},\ldots,\bu_{r-1}^{(t+1)},\bu_{r+1}^{(t)},\ldots, \bu_{R}^{(t)},\bv_{1}^{(t)},\ldots,\bv_{R}^{(t)},\bw_{1}^{(t)},\ldots,\bw_{R}^{(t)},\lambda_{1}^{(t)},\ldots,\lambda_{R}^{(t)}, \btheta^{(t)})$. 
Similar conventions apply when defining  $\bTheta^{(t)}_{-\bv_r}$, $\bTheta^{(t)}_{-\bw_r}$, $\bTheta^{(t)}_{-\lambda_r}$ and $\bTheta^{(t)}_{-\btheta}$.
In Algorithm \ref{alg:AGD}, 
after updating each low-rank component $\bu_r$, $\bv_r$ and $\bw_r$, we project them onto the unit sphere to enforce the identifiability constraint. Finally, the iterative algorithm terminates when either the relative change in the log-likelihood $\ell_d(\bTheta)$ falls below a predefined threshold, or the maximum number of iterations is reached.

The initialization algorithm is discussed in Algorithm S1 of the supplement. 
We determine $R$ using the Bayesian Information Criterion (BIC) \citep{schwarz1978estimating}, a model selection method that balances model fit and complexity and has proven effective for low-rank tensor estimation \citep{sun2019dynamic,lee2020tensor,zhang2023learning}. Specifically, we choose the rank $R$ that minimizes
$$
\text{BIC} = \left((d_1+d_2+d_3)\times R+R\right)\times \log|\Omega|-2\ell_d(\hat{\bTheta}),
$$
where $\ell_d$ is the loss function in \eqref{eqn:objective}, and $\hat{\bTheta}$ is the estimate of $\bTheta$ under the working rank. In our numerical experiments, this criterion reliably selects the true rank (see results in Section \ref{sec:tuning}).

\section{Theory}\label{sec:theo} 
In this section, we first establish a non-asymptotic error bound for the estimates obtained from each iteration of Algorithm~\ref{alg:AGD}, and then develop a statistical inference procedure to test whether observation probabilities depend on underlying tensor values. For ease of exposition, we focus on the setting
$g_{\btheta}(\cX_{ijk}) = \text{logit}^{-1} (b_0 + b_1 \cX_{ijk})$, where $\btheta=(b_0,b_1)$. 
We use superscript $\ast$ to denote the true value of the corresponding parameter (e.g., $\cX^\ast$). 
To further simplify notation, we assume $d_1, d_2, d_3=\Theta(d)$. 

\subsection{Non-asymptotic error bound}
To measure the estimation error, we consider the following error metric:
\begin{equation}
\textrm{D}(\bTheta,\bTheta^\ast) =
\max_{r\in[R]}\left\{
\|\bu_{r}-\bu_{r}^\ast\|_2,\;
\|\bv_{r}-\bv_{r}^\ast\|_2,\;
\|\bw_{r}-\bw_{r}^\ast\|_2,\;
\frac{|\lambda_r-\lambda_r^\ast|}{|\lambda_r^\ast|},\frac{|b_0-b_0^\ast|}{|b_0^\ast|},\frac{|b_1-b_1^\ast|}{|b_1^\ast|}
\right\}.
\end{equation}
This definition ensures that errors from estimating different parameters are comparable in scale. 
If $b_0^\ast=0$ ($b_1^\ast=0$), we replace $\frac{|b_0-b_0^\ast|}{|b_0^\ast|}$ ($\frac{|b_1-b_1^\ast|}{|b_1^\ast|}$) in $\textrm{D}(\bTheta,\bTheta^\ast)$ with $|b_0-b_0^\ast|$ ($|b_1-b_1^\ast|$). The main theoretical results in this section still hold.
We first introduce several regularity conditions.
\begin{cond}\label{con1}
Assuming the following conditions hold.
\begin{enumerate}
\item[(a)] Entries in $\cY$ are independent and sub-Gaussian with sub-Gaussian norms, defined as $\|\cY_{ijk}\|_{\psi_2}=\sup_{d\ge1}d^{-1/2}(E|\cY_{ijk}|^d)^{1/d}$, bounded above by some positive constant. 

\item[(b)] The true tensor $\cX^\ast$ satisfies \eqref{eqn:CP} and the decomposition is unique up to a permutation. Let $\lambda_{\min}=\min_r\lambda_r^\ast$, $\lambda_{\max}=\max_r\lambda_r^\ast$ and assume $\lambda_{\min}=\Theta(\lambda_{\max})$. 
\item[(c)] The unit-norm vectors $\bu_{r}^\ast,\bv_r^\ast,\bw_r^\ast$ are $\mu$-mass vectors, $r\in[R]$. That is, $\max_{i}|u_{ri}|\leq\mu/\sqrt{d_1},\max_{j}|v_{rj}|\leq \mu/\sqrt{d_2}$ and $\max_{k}|w_{rk}|\leq \mu/\sqrt{d_3}$ for some positive constant $\mu$. 
\end{enumerate} 
\end{cond}
Assumption \ref{con1}(a) is standard in the theoretical analyses of tensor models \citep{xia2021statistically,cai2019nonconvex,hu2022generalized,zhou2023partially}, and includes common distributions such as Gaussian, Bernoulli and Binomial.
Assumption \ref{con1}(b) is common in CP decomposition based tensor analysis \citep{sun2019dynamic,cai2024jointly}. Assumption \ref{con1}(c) is to ensure that the mass of the tensor does not concentrate on only a few entries. This condition is commonly employed in tensor completion \citep{lee2020tensor,zhou2023partially}. 

Next, we state conditions on the missing probabilities $\cP=(\cP_{ijk})\in \mathbb{R}^{d_1\times d_2\times d_3}$. Unlike existing results that impose a uniform bound on missing probabilities, our requirement is expressed in terms of slice averages of $\cP$, as illustrated in Figure~\ref{fig1}, which is considerably more relaxed. 
Recalling $\cP_{ijk}=\text{logit}^{-1}(b_0^\ast + b_1^\ast \cX_{ijk}^\ast)$, define $\bar{p}$ and $\bar{q}$ as
$$
\bar p = \min\left\{ 
\min_i \frac{1}{d_2 d_3}\sum_{j,k} \cP_{ijk},\;
\min_j \frac{1}{d_1 d_3}\sum_{i,k} \cP_{ijk},\;
\min_k \frac{1}{d_1 d_2}\sum_{i,j} \cP_{ijk}
\right\}.
$$
$$
\bar q = \min\left\{ 
\min_i \frac{1}{d_2 d_3}\sum_{j,k} \cP_{ijk}(1-\cP_{ijk}),\;
\min_j \frac{1}{d_1 d_3}\sum_{i,k} \cP_{ijk}(1-\cP_{ijk}),\;
\min_k \frac{1}{d_1 d_2}\sum_{i,j} \cP_{ijk}(1-\cP_{ijk})
\right\}.
$$
We write  
$\psi_{\max}' = \max_{i,j,k} \left| \psi'(\cX_{ijk}^\ast) \right|$,
$\psi_{\min}'' = \min_{i,j,k} \left| \psi''(\cX_{ijk}^\ast) \right|$,
$\psi_{\max}'' = \max_{i,j,k} \left| \psi''(\cX_{ijk}^\ast) \right|$, 
where $\psi(\cdot)$ denotes the cumulant function in \eqref{eqn:exponential_family} (e.g., $\psi(x)=x^2/2$ for Gaussian data).


\begin{figure}[t!]
\centering
\includegraphics[scale=0.525]{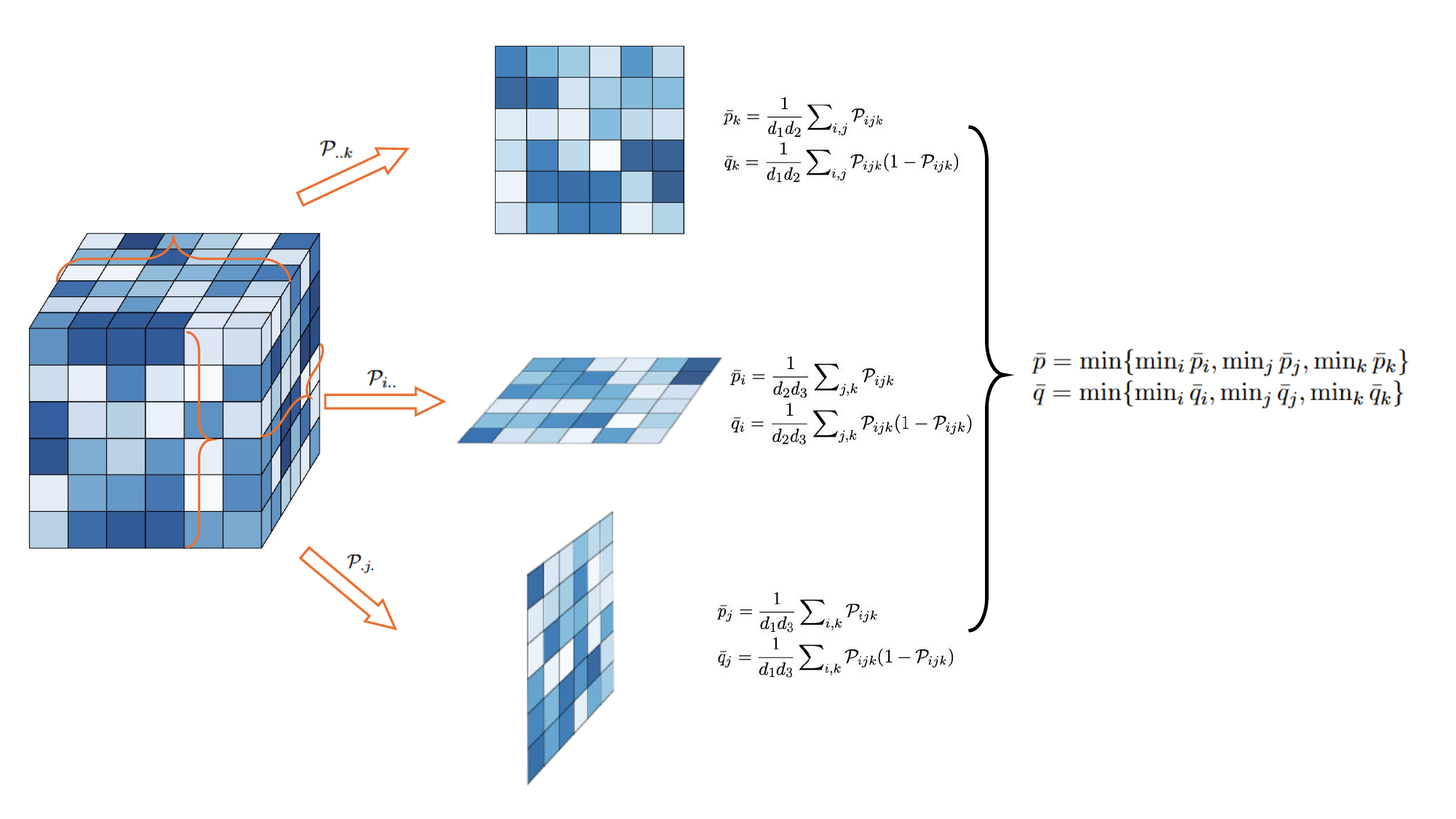}
\caption{Illustration of definitions for $\bar p$ and $\bar q$.} \label{fig1}
\end{figure}

\begin{cond}\label{con2}
Assume that
\begin{equation*}
\bar p\geq C_0\max\left\{\frac{(\psi_{max}'')^2\log^4(d)}{(\psi_{min}'')^2 d^{3/2}},\frac{\{(\psi_{max}')^2+1\}^2\log(d)d}{(\psi_{min}'')^2\lambda_{\min}^2}\right\},
\end{equation*}
\begin{equation*}
\bar q\leq C'_0\bar p(\psi_{\min}'')^2 \text{  or  }
\bar q\geq C_0\frac{\{(\psi_{max}')^2+1\}\log(d)d}{\lambda_{\min}^2}
\end{equation*}
for some positive constants $C_0$ and $C'_0$. When $b_1^\ast\neq 0$, we assume that $b_1^\ast=\Theta(1)$ and $b_0^\ast=\Theta(R\lambda_{\min}/\sqrt{d^3})$. 
\end{cond}
Condition 2 differs in an important way from prior work, which either requires all observation probabilities to be bounded away from 0 and 1 by a constant \citep{ma2019missing,yang2021tenips,li2024pairwise,mao2024mixed} or assumes they are known \citep{chao2021hosvd}. 
In comparison, we allow much greater heterogeneity, including probabilities arbitrarily close to 0 or 1, provided slice-level averages are bounded. 
For tensor completion with Gaussian data and uniform missingness, i.e., $\cP_{ijk}=p$ for all $i,j,k$, Condition \ref{con2} reduces to $b_1^\ast=0$ and 
$p=\frac{1}{1+\exp(-b_0^\ast)}\geq C_0\frac{\log^4(d)}{ d^{3/2}}$. 
This matches with conditions established in the existing literature on tensor completion with uniform missing \citep{cai2019nonconvex,zhou2023partially}. 
For tensor completion with Gaussian data and MNAR missing mechanism, Condition \ref{con2} simplifies to $b_1^\ast=\Theta(1)$, $b_0^\ast=\Theta(R\lambda_{\min}/\sqrt{d^3})$ 
and 
$\bar p\geq C_0\frac{\log^4(d)}{ d^{3/2}}$.  
The conditions on $b_0^\ast$ and $b_1^\ast$ ensure that the observation probabilities are not dominated by either $b_0^\ast$ or $b_1^\ast\cX_{ijk}^\ast$; otherwise, it would be difficult to accurately estimate $b_0^\ast$ or $b_1^\ast$.

\begin{cond}\label{con3}
Suppose $\lambda_{\min}\sqrt{\bar p}\geq C_1\frac{\psi_{\max}''}{(\psi_{\min}'')^2}d^{3/4}\log^2(d)$ for some positive constant $C_1$.
\end{cond}
This condition establishes a lower bound on the signal to noise ratio. Under Gaussian data and uniform missingness, Condition 3 simplifies to $\lambda_{\min}\geq C_1\frac{d^{3/4}\log^2(d)}{\sqrt{p}}$, which matches with the condition in \cite{cai2019nonconvex}. 

We first present the result for the rank-one case ($R=1$) and then extend it to the general rank-$R$ setting. 
\begin{theorem}\label{thm1}
Suppose Conditions 1-3 hold and $R=1$. Assume $\bTheta^{(0)}\in\mathbb{B}_{\frac{1}{2}}(\bTheta^\ast)$, and $ u_{i}^{(0)},v_{j}^{(0)},w_{k}^{(0)}=O(1/\sqrt{d}), i,j,k \in [d]$, the estimator from the $t$-th iteration of Algorithm~\ref{alg:AGD} satisfies that, with probability at least $1-o(1)$,
\begin{equation*}
\textrm{D}(\bTheta^{(t)},\bTheta^\ast)\leq \rho^t \textrm{D}(\bTheta^{(0)},\bTheta^\ast)+C_2\frac{\left(\psi_{\max}'+1\right)\sqrt{\log(d)d}}{(\psi_{\min}''\sqrt{\bar p}+\sqrt{\bar q})\lambda^\ast},
\end{equation*}
where the constant $0<\rho<1$ is defined in (S73) and $C_2$ is some positive constant. 
\end{theorem}
The non-asymptotic error bound in Theorem~\ref{thm1} includes two terms, the first one is the computational error and it decreases geometrically with the iteration number $t$. The second term is the statistical error, which is related to the model parameters and is independent of $t$. Correspondingly, when the iteration $t\geq \log(C_2\frac{\left(\psi_{\max}'+1\right)\sqrt{\log(d)d}}{(\psi_{\min}''\sqrt{\bar p}+\sqrt{\bar q})\lambda^\ast})/\log(\rho)$, the computational error is dominated by the statistical error and the estimator in Algorithm~\ref{alg:AGD} converges geometrically to a neighborhood that is within statistical error of the true unknown parameter. In the special case of the uniform missing where $\cP_{ijk}$ are all equal, our statistical error reduces to that in the literature \citep{cai2019nonconvex,lee2020tensor}. More generally, unlike prior work on non-uniform missing \citep{ma2019missing,li2024pairwise,mao2024mixed}, which assumes missing probabilities are bounded away from 0 and 1 by a constant, 
our result allows heterogeneous $\cP_{ijk}$, including some approaching $0$ and others near $1$.

The theoretical analysis for $R>1$ is more challenging because components from different ranks are not orthogonal, which can lead to strong correlations across factors. To control these correlations, we impose the following incoherence condition.
\begin{cond}\label{con4}
Define the incoherence parameter as $\xi=\max\limits_{r'\neq r}\{\langle \bu_{r}^\ast,\bu_{ r'}^\ast\rangle,\langle \bv_{ r}^\ast,\bv_{ r'}^\ast\rangle,\langle \bw_{ r}^\ast,\bw_{r'}^\ast\rangle\}$ and assume that 
\begin{equation*}
\xi\leq\frac{C_3\lambda_{\min}^2}{\lambda_{\max}^2R^2}
\end{equation*}
for some positive constant $C_3$.
\end{cond}
This condition ensures that components across different ranks are not highly correlated. When $\xi=0$, the components are perfectly orthogonal. Moreover,  when rank $R$ increases, the upper bound on $\xi$ becomes tighter, reflecting the need for stronger separation among components. Similar incoherence conditions have been adopted in \cite{sun2019dynamic,zhou2023partially,cai2024jointly}.

\begin{theorem}\label{thm2}
Suppose Conditions~\ref{con1}-\ref{con4} hold. Given $\bTheta^{(0)}\in\mathbb{B}_{\frac{1}{2}}(\bTheta^\ast)$, and $ u_{ri}^{(0)},v_{rj}^{(0)},w_{rk}^{(0)}=O(1/\sqrt{d})$, 
the estimator from the $t$-th iteration of Algorithm~\ref{alg:AGD} satisfies, with probability at least $1-o(1)$,
\begin{equation*}
\textrm{D}(\bTheta^{(t)},\bTheta^\ast)\leq \rho_R^t \textrm{D}(\bTheta^{(0)},\bTheta^\ast)+C_4\frac{\left(\psi_{\max}'+1\right)\lambda_{\max}\sqrt{\log(d)d}}{(\psi_{\min}''\sqrt{\bar p}+\sqrt{\bar q})\lambda_{\min}^2},
\end{equation*}
where the constant $0<\rho_R<1$ is defined in (S118) and $C_4$ is some positive constant. 
\end{theorem}
Similar to Theorem~\ref{thm1}, the non-asymptotic error bound consists of two terms: a computational error and a statistical error. When the number of iterations is sufficiently large, e.g., $t\geq \log(C_4\frac{\left(\psi_{\max}'+1\right)\lambda_{\max}\sqrt{\log(d)d}}{(\psi_{\min}''\sqrt{\bar p}+\sqrt{\bar q})\lambda_{\min}^2})/\log(\rho_R)$, the computational error becomes dominated by the statistical error.

\subsection{Key technical challenges} 

We highlight the main challenges in our proof. 
First, both the tensor data model and observation probability model in \eqref{eqn:joint_likelihood} are generalized linear models. 
As a result, the likelihood-based parameter updates at each iteration do not admit closed-form solutions. Our analysis therefore require a delicate control of the gradient and Hessian of the joint likelihood function, which is more involved than the squared-loss setting typically studied in the tensor completion literature \citep{cai2019nonconvex, sun2019dynamic, yang2021tenips, zhou2023partially}.

Second, by performing joint estimation, we allow the observation probabilities to be arbitrarily close to 0 or 1, thereby greatly relaxing the conditions imposed in prior work. This, however, introduces significant challenges for our theoretical analysis, since it is no longer adequate to impose uniform upper and lower bounds on the probabilities, as is commonly done in the literature \citep{ma2019missing,li2024pairwise,mao2024mixed}.
To tackle this challenge, we develop a new proof strategy that refines the standard strong concavity argument by analyzing individual coordinates of the low-rank components, rather than working at the vector level. This coordinate-wise approach provides a sharper control of the statistical error, especially in the presence of highly heterogeneous $\cP_{ijk}$.

We illustrate a few other challenges through the following example.
Consider $R = 1$ and focus on the estimation of $u_i^\ast$, the $i$th entry of $\bu^\ast$. We begin by establishing a strong concavity condition for $u_i$, a critical condition in nonconvex optimization analysis, conditional on the remaining parameters $\bar\bTheta_{-u_i}=(\bar u_1,\ldots,\bar u_{i-1},\bar u_{i+1},\ldots,\bar u_{d_1},\bar\bv,\bar\bw,\bar\lambda,\bar\btheta)$, which include all other model parameters, with some already updated while others yet to be updated. Specifically, for any $u_i'$, we have
\[
-\frac{1}{2}(u_i' - u_i^\ast)^2 \nabla_{u_i}^2 \ell_d(u_i'', \bar\bTheta_{-u_i}) = \ell_d(u_i^\ast, \bar\bTheta_{-u_i}) - \ell_d(u_i', \bar\bTheta_{-u_i}) + \left\langle \nabla_{u_i} \ell_d(u_i^\ast, \bar\bTheta_{-u_i}), u_i' - u_i^\ast \right\rangle,
\]
where $u_i''$ lies between $u_i'$ and $u_i^\ast$. Since $u_i'$ maximizes $\ell_d(\cdot, \bar\bTheta_{-u_i})$, the term $\ell_d(u_i^\ast, \bar\bTheta_{-u_i}) - \ell_d(u_i', \bar\bTheta_{-u_i})$ on the right-hand side is non-positive, leaving us with two main tasks: obtaining a lower bound for $-\nabla_{u_i}^2 \ell_d(u_i'', \bar\bTheta_{-u_i})$ and an upper bound for $\left\langle \nabla_{u_i} \ell_d(u_i^\ast, \bar\bTheta_{-u_i}), u_i' - u_i^\ast \right\rangle$. The Hessian term $-\nabla_{u_i}^2 \ell_d(u_i'', \bar\bTheta_{-u_i})$ expands as
\begin{equation}\label{strongconexp}
 \frac{\bar\lambda^2}{\phi_0} \sum_{j,k} \bar v_j^2 \bar w_k^2 \cD_{ijk} \psi''(\bar\lambda u_i'' \bar v_j \bar w_k)
+ \bar b_1^2 \bar\lambda^2 \sum_{j,k} \bar v_j^2 \bar w_k^2 \sigma_{ijk}(u_i'', \bar\bTheta_{-u_i}) \left[1 - \sigma_{ijk}(u_i'', \bar\bTheta_{-u_i})\right],
\end{equation}
where $\sigma_{ijk}(\bTheta)=\frac{1}{1+\exp(-b_0-b_1\cX_{ijk})}$.
When $\cP_{ijk}$'s can be arbitrarily close 0 or 1, 
the population quantity $\sum_{j,k} \bar v_j^2 \bar w_k^2\mathbb{E}(\cD_{ijk}) \psi''(\cdot)=\sum_{j,k} \bar v_j^2 \bar w_k^2\cP_{ijk} \psi''(\cdot)$ may be offset by the fluctuation term $\sum_{j,k} \bar v_j^2 \bar w_k^2 (\cD_{ijk} - \cP_{ijk}) \psi''(\cdot)$, requiring more refined arguments based on average probabilities (e.g., $\bar p$) rather than uniform bounds of $\cP_{ijk}$'s. 
Another difficulty is that even when $\bar\bTheta$ is close to $\bTheta^\ast$, the logistic transformation can map small deviation in $\hat\cX_{ijk}$ into large differences in observation probabilities. Consider, for example, $b_0 = \log d$, $b_1 = 1$, $\cX_{ijk} = \log d$ and  $\hat{\cX}_{ijk} = (1+\eta)\log d$, then $\cP_{ijk} = \frac{1}{1+d^2}$ but $\hat\cP_{ijk} = \frac{1}{1+d^{2+\eta}}$. This nonlinearity amplifies small estimation errors and requires a careful control in the theoretical analysis.
Next, the term $\langle \nabla_{u_i} \ell_d(u_i^\ast, \bar\bTheta_{-u_i}), u_i' - u_i^\ast \rangle$ can be bounded as
\begin{equation*}
\begin{aligned}
&\left| \left\langle \nabla_{u_i} \ell_d(u_i^\ast, \bar\bTheta_{-u_i}), u_i' - u_i^\ast \right\rangle \right| \\
\leq& \left| \nabla_{u_i} \ell_d(u_i^\ast, \bTheta_{-u_i}^\ast) \right| \left| u_i' - u_i^\ast \right| 
+ \left| \nabla_{u_i} \ell_d(u_i^\ast, \bar\bTheta_{-u_i}) - \nabla_{u_i} \ell_d(u_i^\ast, \bTheta_{-u_i}^\ast) \right| \left| u_i' - u_i^\ast \right|.
\end{aligned}
\end{equation*}
The second term captures the effect of replacing $\bTheta_{-u_i}^\ast$ with $\bar\bTheta_{-u_i}$ in the sample-level gradient and is particularly challenging to control due to several reasons:  
(i) it requires sharp concentration inequalities to bound the differences between sample-level gradient and population-level gradient; 
(ii) the bound must hold uniformly over all admissible $\bar\bTheta$ satisfying the regularity conditions; and  
(iii) $\bar\bTheta_{-u_i}$ contains mixed types of parameters (vectors and scalars), so standard Lipschitz arguments with respect to $\bTheta_{-u_i}$ are not directly applicable.  
To address these difficulties, we exploit 
employ covering number arguments and derive uniform concentration bounds for our setting.

\subsection{Hypothesis testing on $b_1$}

In this section, we study hypothesis testing for $b_1$. Formally, we test
\begin{equation}
\label{eqn:test}
H_0:\, b_1^\ast=0 \text{ vs. } H_a: \, b_1^\ast\neq 0.
\end{equation}
Under the null, the observation probability does not depend on the underlying tensor entries $\cX_{ijk}^\ast$'s. In this case, missingness is MCAR since $\cP_{ijk}=\text{logit}^{-1}(b_0^\ast)$. Under the alternative, the observation probability depends on $\cX_{ijk}$, leading to a MNAR mechanism. 

Testing the hypothesis in \eqref{eqn:test} is challenging, as $\bTheta$ is estimated from a non-convex optimization problem via an alternative maximization algorithm. If $\cX$ were known, then the test in \eqref{eqn:test} would reduce to a standard inference problem in logistic regression, with $\cD_{ijk}$ as the response and $\cX_{ijk}$ as the predictor. However, in practice, $\cX$ must be estimated from $\cY_{\mathrm{obs}}$ and $\cD$, leading to two issues: (i) the estimated covariates $\hat\cX_{ijk}$ are statistically dependent on $\cD_{ijk}$, and (ii) $\hat\cX$ contains estimation error, which must be carefully controlled to ensure valid inference.

We consider a sample splitting strategy that divides the data into two subsamples: one used to estimate $\cX$ and the other to test $b_1$. This approach ensures that $\hat\cX$, estimated from the estimation subsample, is independent of $\cD$ from the testing subsample. Given $\cX$, testing $b_1$ reduces to inference in a two-parameter logistic regression, which requires far fewer samples than that needed to accurate estimating $\cX$. Consequently, the testing subsample can be taken negligible in size relative to the estimation subsample, so that almost all data contribute to accurately estimate the underlying tensor, while retaining enough data for valid inference on $b_1$.

Our proposed testing procedure consists of three steps. First, we randomly split all indexes in $\{(i,j,k):i\in[d_1],j\in[d_2],k\in[d_3]\}$
into two disjoint subsets, denoted as $\mathcal{A}_1$ and $\mathcal{A}_2$.
We use $\mathcal{A}_1$ for estimating $\cX$ and $\mathcal{A}_2$ for testing $b_1$. 
The size of the testing set $\mathcal{A}_2$ satisfies
\[
1 \prec \sqrt{|\mathcal{A}_2|} \prec \frac{(\psi_{\min}'' \sqrt{\bar{p}} + \sqrt{\bar{q}})d \lambda_{\min}^2}{\left( \psi_{\max}' + 1 \right) \lambda_{\max}^2 \sqrt{ \log(d)}},
\]
which ensures a sufficient number of entries are retained for accurate estimation. 
Under this condition, the expected number of observed entries in $\mathcal{A}_1$ is $(1-o(1))|\Omega|$. 

Next, we apply Algorithm~\ref{alg:AGD} to entries in $\mathcal{A}_1$. The log-likelihood function is written as
\begin{equation*}
\begin{aligned}
\ell_{\cA_1}(\bTheta)=&\sum_{i,j,k \in \cA_1}\cD_{ijk}\log (f(\cY_{ijk}))+\sum_{i,j,k \in \cA_1}\cD_{ijk}(b_0+b_1\cX_{ijk})-\sum_{i,j,k \in \cA_1}\log(1+\exp(b_0+b_1\cX_{ijk})).
\end{aligned}
\end{equation*}
After obtaining the estimated tensor parameters from $\mathcal{A}_1$, denoted as $\hat{\bTheta}^{\mathcal{A}_1}$, we compute the estimated tensor $\hat{\cX}_{ijk}^{\mathcal{A}_1}$. 
We then fit a logistic regression model to the entries in $\mathcal{A}_2$, using $\hat{\cX}_{ijk}^{\mathcal{A}_1}$ as the predictor. 
Denote the estimated parameters by $\hat{\btheta}^{\mathcal{A}_2}=(\hat b_0^{\mathcal{A}_2},\hat b_1^{\mathcal{A}_2})$. To construct the test statistic, we employ results from the following theorem.
\begin{theorem}\label{thm3}
Assume that all conditions in Theorems~\ref{thm1} and \ref{thm2} hold. If $\frac{\left(\psi_{\max}'+1\right)\lambda_{\max}^2\sqrt{|\mathcal{A}_2|\log(d)}}{(\psi_{\min}''\sqrt{\bar p}+\sqrt{\bar q})d\lambda_{\min}^2}=o(1)$, then
\begin{equation*}
\left\{-\nabla^2_{\btheta}\ell_{\mathcal{A}_2}(\hat{\btheta}^{\mathcal{A}_2})\right\}^{1/2}(\hat{\btheta}^{\mathcal{A}_2}-\btheta^\ast)\stackrel{\mathcal{D}}{\rightarrow} N(\bm{0},\bm{I}),
\end{equation*}
where 
$\nabla^2_{\btheta}\ell_{\mathcal{A}_2}(\hat{\btheta}^{\mathcal{A}_2})=-\sum\limits_{(i,j,k) \in \cA_2}\frac{\exp(-\hat{b}_0^{\mathcal{A}_2}-\hat{b}_1^{\mathcal{A}_2}\hat\cX_{ijk}^{\mathcal{A}_1}))}{\{1+\exp(-\hat{b}_0^{\mathcal{A}_2}-\hat{b}_1^{\mathcal{A}_2}\hat\cX_{ijk}^{\mathcal{A}_1})\}^2}\left(\begin{matrix}
1 & \hat\cX_{ijk}^{\mathcal{A}_1}\\
\hat\cX_{ijk}^{\mathcal{A}_1} & (\hat\cX_{ijk}^{\mathcal{A}_1})^2\\
\end{matrix}\right).$
\end{theorem}
This theorem establishes the asymptotic distribution of $\hat {\btheta}^{\mathcal{A}_2}$. 
Note that Theorems 1 and 2 require $\frac{\left(\psi_{\max}'+1\right)\lambda_{\max}^2\sqrt{\log(d)}}{\left\{\psi_{\min}''\sqrt{\bar p}+\sqrt{\bar q}\right\}d\lambda_{\min}^2}=o(1)$ for estimation consistency of $\cX$.
Since $|\mathcal{A}_2|\rightarrow\infty$, the condition in Theorem 3 is stronger than $\frac{\left(\psi_{\max}'+1\right)\lambda_{\max}^2\sqrt{\log(d)}}{\left\{\psi_{\min}''\sqrt{\bar p}+\sqrt{\bar q}\right\}d\lambda_{\min}^2}=o(1)$, due to the additional requirement that the estimation error of $\hat\cX$ be sufficiently small in order to achieve valid asymptotic inference.


We construct a test statistic for $b_1$ as
$$
Z =\{-\nabla^2_{\btheta}\ell_{\mathcal{A}_2}(\hat\btheta^{\mathcal{A}_2})\}^{1/2}_{2,2}\,\hat b_1^{\mathcal{A}_2},
$$
where $\{-\nabla^2_{\btheta}\ell_{\mathcal{A}_2}(\hat\btheta^{\mathcal{A}_2})\}^{1/2}_{2,2}$ denotes the $(2,2)$ entry of the observed information matrix. 
By Theorem~\ref{thm3}, $Z$ converges in distribution to the standard normal distribution. This gives, for example, $p$-values for testing $H_0:b_1^\ast=0$ and confidence intervals of the form
$$
\hat b_1^{\mathcal{A}_2} \;\pm\; z_{\alpha/2}\,\{-\nabla^2_{\btheta}\ell_{\mathcal{A}_2}(\hat\btheta^{\mathcal{A}_2})\}_{2,2}^{-1/2}.
$$

\section{Simulation}\label{sec:sim}
In this section, we evaluate the performance of the proposed model on continuous, binary, and count data, and compare it with alternative methods. For continuous and count data, we quantify performance using the Relative Mean Square Error (RMSE) evaluated on the missing entries, defined as
$$
\text{RMSE} = \frac{\|(1-\cD) * \hat{\mathcal{X}} - (1-\cD) * \mathcal{X}^\ast\|_F}{\|(1-\cD) * \mathcal{X}^\ast\|_F},
$$
where \(\hat{\mathcal{X}}\) represents the tensor estimated by different methods. For binary data, we assess accuracy using the AUC ROC evaluated on the missing entries.
We also include the RMSE and AUC ROC evaluated on all entries, including the observed ones, in Section S5 of the supplement.

\subsection{Gaussian data}\label{sec:sim_con}
To generate $\cX^\ast$, we first generate entries in $\bu_{r}^\ast,\bv_{r}^\ast$ and $\bw_{r}^\ast$ independently from $\mathcal{N}(0.5,1)$ and then scale $\bu_{r}^\ast,\bv_{r}^\ast$ and $\bw_{r}^\ast$ to be unit-length vectors. We let $\lambda_{r}^\ast=cd^{3/2}$ with $c=0.2, 0.4, 0.6,0.8,1$, representing varying levels of signal strength. We generate $\cY_{ijk}$ from $N(\cX_{ijk},1)$.  
The entries in missingness mask $\cD$ are generated from the Bernoulli distribution with $\operatorname{logit}\left(\cP_{i j k}\right)=b_0^\ast+b_1^\ast \mathcal{X}_{i j k}^\ast$, where $b_1^\ast=2$. We vary $b_0^\ast$ in $\{-1,-0.5,0,0.5,1,1.5,2\}$, corresponding to observations ratio of 0.36, 0.44, 0.52, 0.60, 0.68, 0.76 and 0.84, respectively. We set $d_1=d_2=d_3=50$ and $R=3$.

We compare our method, referred to as \textbf{GTC-MNAR}, with four existing tensor completion methods: (1) \textbf{CP-ALS}, the standard alternating least squares method \citep{kolda2009tensor}; (2) \textbf{GCP}, the generalized CP decomposition method \citep{hong2020generalized}; (3) \textbf{Online}, the online tensor completion method of \cite{cai2023online} and \cite{wen2023online}, where entires are observed sequentially. For a fair comparison, the online algorithm terminates after $|\Omega|$ steps to ensure that the number of observed entries matches that of our method; and (4) \textbf{TenIPS}, the inverse propensity weighing method proposed by \cite{yang2021tenips}. 
Performance is evaluated under each simulation setting over 50 data replicates. 

\begin{figure}[ht!]
\centering
\includegraphics[width=0.9\linewidth]{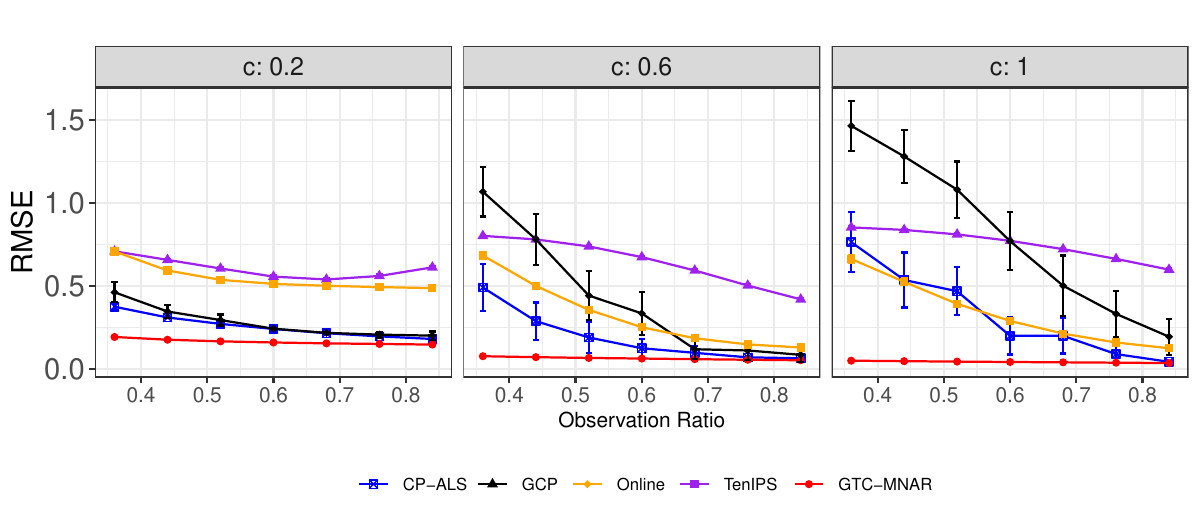}
\caption{Mean RMSE with 95\% confidence intervals in the Gaussian case. 
} \label{fig:continuous}
\end{figure}

Figure \ref{fig:continuous} presents the mean RMSE and 95\% confidence intervals for missing entries across different settings. Additional results for $c=0.4$ and $c=0.8$ are reported in Section S5 of the supplement. It is seen that GTC-MNAR consistently outperform other methods,
and its suprior performance is most noticeable at lower observation ratios (below 60\%). 
The performance of TenIPS is sensitive to small observation probabilities as the method uses inverse propensity score weighing. 

As $c$ increases, corresponding to a higher signal-to-noise ratio, the RMSE of our method decreases. In contrast, the RMSE of the other methods tends to increase. This pattern occurs because larger values of $c$ yields larger entries in $\cX^\ast$, which strengthens the effect of the MNAR mechanism. 
While our method is able to benefit from the stronger signal, the alternative methods are more affected by the non-random missing pattern.

\subsection{Bernoulli data}\label{sec:sim_bi}
We generate $\cX^\ast$ and $\cD$ as in Section \ref{sec:sim_con} and generate $\cY_{ijk}$ from $\text{Bernoulli}(\sigma(\cX^\ast))$. 
We compare our proposed method, \textbf{GTC-MNAR}, with four alternative methods: (1) \textbf{GCP} by \cite{hong2020generalized}, (2) \textbf{NonparaT}, the nonparametric tensor completion via sign series by \cite{lee2021beyond}, (3) \textbf{Ordinal}, the ordinal tensor completion method that uses a multi-linear cumulative link model \citep{lee2020tensor}, and (4) \textbf{ENTED}, a nonparametric tensor decomposition method proposed by \cite{tao2024efficient}. 
Performance is evaluated under each simulation setting over 50 data replicates.

\begin{figure}[ht!]
		\centering
		\includegraphics[width=0.9\linewidth]{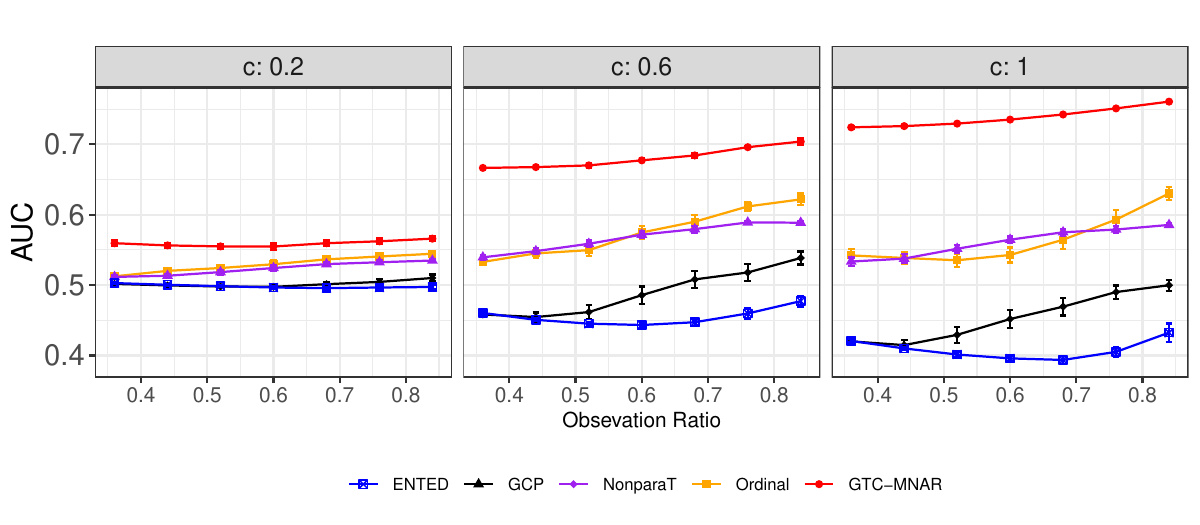}
	\caption{Mean AUC with 95\% confidence intervals in the Bernoulli case. 
 } \label{fig:binary_pre}
\end{figure}

Figure \ref{fig:binary_pre} presents the mean AUC ROC and 95\% confidence intervals for missing entries
across different settings.  Additional results for $c = 0.4$ and $c = 0.8$ are reported in Section S5 of the supplement. 
It is seen that GTC-MNAR consistently outperform other methods, and the performance improves with $b_0$ and $c$. 
Notably, as the signal level $c$ grows, our method leverages the stronger signal to further improve AUC, while other methods, designed mainly for MCAR, show weaker performance under the strengthened effects of MNAR.

\subsection{Poisson data}\label{sec:sim_count}
To generate $\cX^\ast$, we first generate entries in $\bu_{r}^\ast,\bv_{r}^\ast$ and $\bw_{r}^\ast$ independently from $\text{Unif}(0,1)$, take the logarithm and then scale $\bu_{r}^\ast,\bv_{r}^\ast$ and $\bw_{r}^\ast$ to be unit-length vectors. We let $\lambda_{r}^\ast=cd^{3/2}$ with $c=0.3,0.4,0.5$, representing varying levels of signal strength. 
We generate $\cY_{ijk}$ from $\text{Poisson}(\exp(\cX_{ijk}^\ast))$.  
The entries in missingness mask $\cD$ are generated from the Bernoulli distribution with $\operatorname{logit}\left(\cP_{i j k}\right)=b_0^\ast+b_1^\ast \mathcal{X}_{i j k}^\ast$, where $b_0^\ast=0.5$. We vary $b_1^\ast$ in $\{-2, -1, 0, 1, 2, 3,4\}$, corresponding to observations ratio of 0.77, 0.7, 0.63, 0.56, 0.49, 0.42, and 0.35, respectively. We set $d_1=d_2=d_3=d=50$ and $R=3$. 
We compare our proposed method, \textbf{GTC-MNAR}, with four existing tensor methods: (1) \textbf{GCP} by \cite{hong2020generalized}, (2) \textbf{Ordinal} by \cite{lee2020tensor}, (3) \textbf{ENTED} by \cite{tao2024efficient} and (4) \textbf{MTDF}, a non-linear tensor decomposition method using neural networks \citep{fan2021multi}. 
Performance is evaluated under each simulation setting over 50 data replicates.

Figure \ref{fig:count_pre_logunif} reports the RMSE and 95\% confidence intervals for missing entries
across different settings. It is seen that GTC-MNAR consistently outperform other methods, and its performance improves with $b_1$ and $c$. 
In this evaluation, we also consider $b_1^\ast=0$, corresponding to the MCAR scenario. 
Under this setting, our method still performs better than the other methods, likely due to the fact our method directly uses the Poisson likelihood function as the objective. 

\begin{figure}[t!]
\centering
\includegraphics[width=0.9\linewidth]{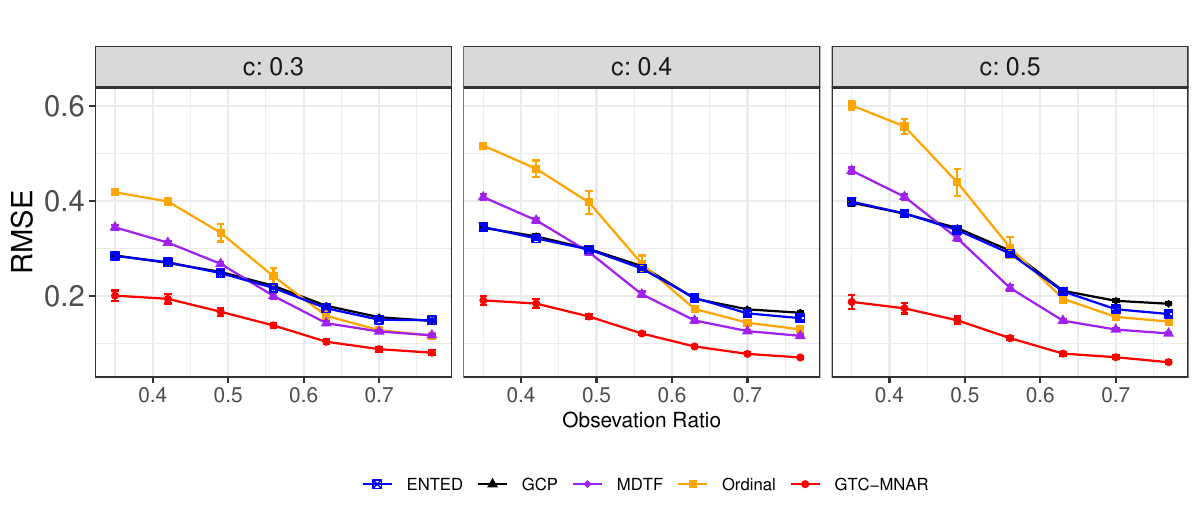}
\caption{Mean RMSE with 95\% confidence intervals in the Poisson case.
} \label{fig:count_pre_logunif}
\end{figure}





\subsection{Parameter tuning for rank}\label{sec:tuning}
We further evaluate our tuning procedure for rank $R$ selection. Due to space limitations, in the main paper we present the tuning results for the Gaussian data (Section \ref{sec:sim_con}) and Bernoulli data ( Section \ref{sec:sim_bi}), while the results for the Poisson case are reported in Section S5 of the supplement. We set $b_0^\ast$ to be -1, 0, 1, and 2, corresponding to observation ratios of approximately 0.36, 0.52, 0.68, and 0.84, respectively. Candidate ranks from 2 to 10 are considered. Figure \ref{fig:tuning_R} shows boxplots of ranks selected by BIC across 50 data replicates. Overall, the selected ranks closely match the true rank $R=3$ in most scenarios, and tuning performance improves when the signal level increases. 




\begin{figure}[ht!]
\centering
\includegraphics[width=0.45\linewidth]{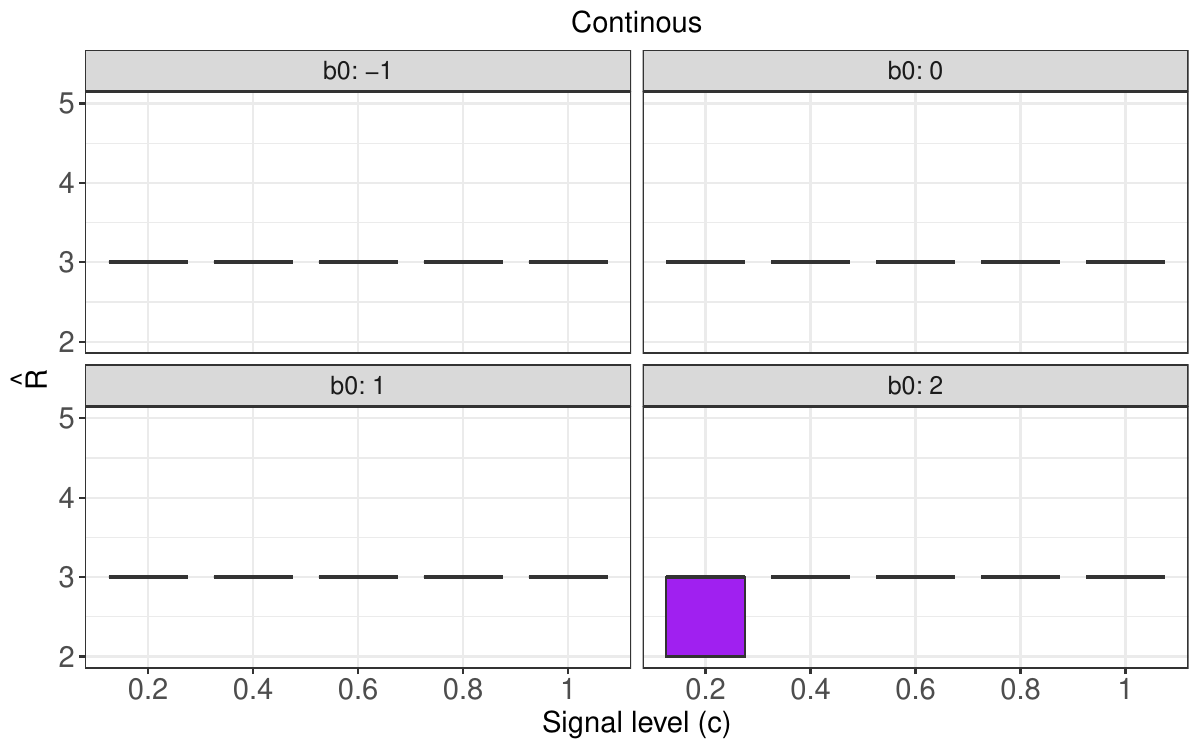}
\centering
\includegraphics[width=0.45\linewidth]{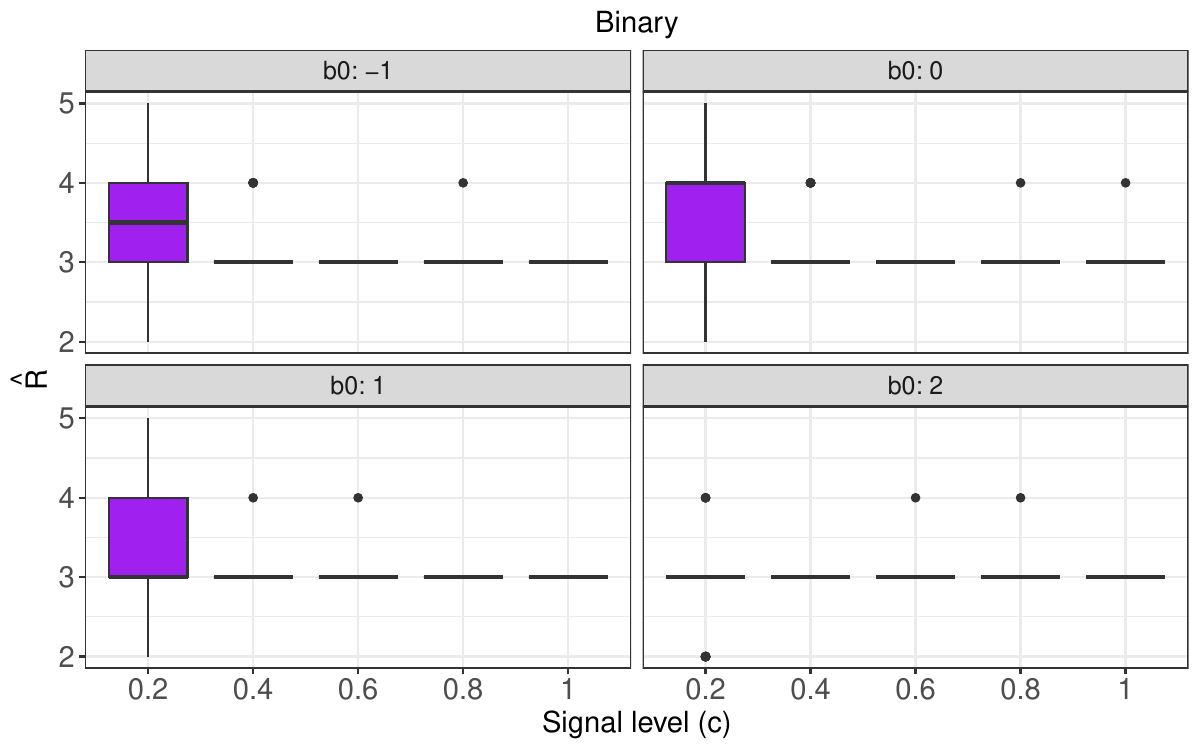}
\caption{Boxplots of selected ranks in Gaussian data (left plot) and Bernoulli data (right plot). The true rank is 3 in all scenarios.
} \label{fig:tuning_R}
\end{figure}

\subsection{Hypothesis testing}
To evaluate the proposed hypothesis testing procedures, we consider the same settings as in Section \ref{sec:sim_con} and Section \ref{sec:sim_bi}, with $c=0.6$ and $b_0^\ast=1$. We set $d_1=d_2=50$, and vary $d_3 \in \{50,100,150\}.$ The results for Poisson data are included in Section S5 of the Supplement. 
We test the null hypothesis \( H_0: b_1^\ast = 0 \) using the test statistics derived from Theorem \ref{thm3}. We set $|\mathcal{A}_2|=500$ and the significance level to 0.05.  
Tables \ref{tab:power_size_continious} and \ref{tab:power_size_binary} summarize the results for Gaussian and Bernoulli data, respectively. The tables show that the empirical size (setting with $b_1^*=0$) is close to the nominal level, and the empirical power (settings with $b_1^*=0.5,2$) is close to 1. 

\begin{table}[!t] 
  \centering  
  \caption{Proportion of rejections under Gaussian data with $b_1^*=0, 0.5,2$. }
  \vspace{0.5em}
  \begin{tabular}{|l|c|c|c|}
  	\hline
  	  &$d_3=50$ & $d_3=100$&$d_3=150$\\
      \hline
        $b_1^*=0$ & 0.052 &0.048&0.048\\
  	\hline
  	$b_1^*=0.5$ & 1 & 1& 1\\
  	\hline
        $b_1^*=2$ & 1 & 1  &1 \\
          \hline

  \end{tabular}
  \label{tab:power_size_continious}
\end{table}

\begin{table}[!t] 
  \centering  
  \caption{Proportion of rejections under Bernoulli data with $b_1^*=0, 0.5,2$.}
  \vspace{0.5em}
  \begin{tabular}{|l|c|c|c|}
  	\hline
  	  &$d_3=50$ & $d_3=100$&$d_3=150$\\
      \hline
       $b_1^*=0$ & 0.055 &0.051&0.048\\
  	\hline
       $b_1^*=0.5 $ & 0.994 & 0.997& 0.999\\
  	\hline
        $b_1^*=2$ & 1 & 1  &1 \\
          \hline

  \end{tabular}
  \label{tab:power_size_binary}
\end{table}

\section{Real data application}\label{sec:real}
In this section, we illustrate our method on two real datasets: a music recommendation dataset and an advertisement rating dataset.

\subsection{InCarMusic data}
The InCarMusic dataset \footnote{\url{https://www.kaggle.com/datasets/stefanogiannini/carskit-contextaware-music}} records ratings of 139 songs (on a scale of 1 to 5) from 42 users across 26 contexts, such as weather, mood, and landscape. Figure \ref{fig:ADS_sum} shows the distribution of observed ratings, which is heavily skewed toward low values.  

\begin{figure}[!ht]
\centering
\includegraphics[width=0.5\linewidth]{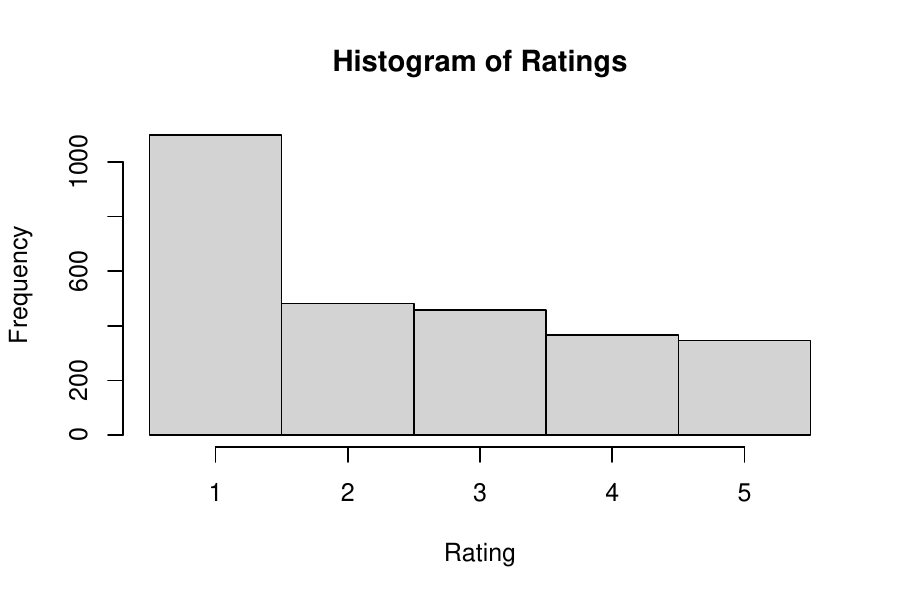}
\caption{The distribution of ratings from the InCarMusic data.
} \label{fig:ADS_sum}
\end{figure}

This dataset forms a tensor of size $42\times 139 \times 26$, with only 2844 entries observed, corresponding to a sparsity level of approximately 2\%. 
Such a high degree of missingness makes modeling multi-class outcomes difficult. To mitigate this, we dichotomized the ratings by treating values of 4 or 5 as positive evaluations (coded as 1) and ratings below 4 as non-positive evaluations (coded as 0). 

We apply the hypothesis testing procedure from Section~\ref{sec:theo} to examine whether the missingness probability depends on the underlying tensor values, testing
$H_0: b_1^\ast=0 \quad \text{vs.} \quad H_a: b_1^\ast\neq 0$. 
The size of $\mathcal{A}_2$ is set to 10\% of $42\times 139 \times 26$.
The test yields a $p$-value of $1.67\times 10^{-43}$, providing overwhelming evidence that $b_1$ differs from zero. 
The 95\% confidence interval of $\hat{b}_1$ is $[-0.136,-0.102]$, suggesting that larger ratings are associated with a higher probability of being missing, consistent with the patterns observed in Figures \ref{fig:ADS_sum}.

Next, we perform edge prediction by splitting 80\% of the observed data into a training set and the remaining 20\% into a testing set. We then use 5-fold cross-validation to compute the mean and standard error of prediction AUC ROC. Table \ref{tab:music_binary} presents the average prediction errors obtained from 5-fold cross-validation. Based on the BIC in Section \ref{sec:model}, our method selected a rank of $R=5$ in three out of the five folds, with one fold selecting \(R = 6\) and another selecting \(R = 4\). 
For a fair comparison, we used the same rank for the other methods within each fold. 
It is seen that our method achieves the best performance.

\begin{table}[!t] 
  \centering  
  \caption{Average AUC ROC in the InCarMusic data with standard errors in parentheses.}
  \vspace{0.5em}
  \begin{tabular}{|c|c|c|c|c|}
  	\hline
  	 GTC-MNAR  &GCP& Ordinal&NonparaT&ENTED\\
  	\hline
  	 $\mathbf{0.702}$ & 0.659 & 0.529 & 0.642 & 0.638\\
          (0.006) & (0.009) & (0.010)  & (0.008) & (0.013)\\
  	\hline
  \end{tabular}
  \label{tab:music_binary}
\end{table}

\subsection{ADS Data}
The ADS dataset\footnote{\url{https://www.kaggle.com/datasets/groffo/ads16-dataset}} \citep{roffo2016personality} contains ratings of 300 advertisements by 120 individuals. Participants rated each advertisement on a scale of 1 to 5 based on their willingness to click. 
The advertisements are divided into three formats, Rich Media, Image, and Text, and span 20 product and service categories. 
This yields a $120 \times 20 \times 3$ binary tensor capturing interactions between participants, product categories, and formats. Each entry records whether a participant is willing to click on advertisements within a given category and format, with average ratings above 4 coded as 1 and below 4 coded as 0.


Since the ADS tensor is fully observed, we introduce missingness under three mechanisms: no dependence ($b_1^\ast=0$), positive dependence ($b_1^\ast>0$), and negative dependence ($b_1^\ast<0$).
For each mechanism, we randomly mask 20\% of the entries as the test set and use the remaining 80\% for training. Based on the BIC criterion, the selected rank was $R=2$ across all five folds, and we set the same rank for competing methods to ensure fairness. 
Table \ref{tab:ADS_binary} reports the average prediction AUC ROC from 5-fold cross-validation with standard errors. Across all missingness mechanisms, our proposed method achieves the best performance. 

\begin{table}[h!] 
  \centering  
  \caption{Average AUC ROC in the ADS data with standard errors in parentheses.}
  \vspace{0.5em}
  \begin{tabular}{|c|c|c|c|c|c|}
  	\hline
  	  &Proposed  &GCP& Ordinal&NonparaT&ENTED\\
  	\hline
       No & $\mathbf{0.869}$ & 0.839 & 0.826 & 0.639 & 0.808\\
         & (0.009) & (0.006) & (0.011)  & (0.011) & (0.011)\\
        \hline
       Positive & $\mathbf{0.921}$ & 0.894 & 0.899 & 0.639 & 0.867\\
  	 & (0.012) & (0.013) & (0.014) & (0.014) & (0.010)\\
  	\hline
  	Negative & $\mathbf{0.820}$& 0.799& 0.756& 0.567& 0.733 \\
  	
        & (0.005)& (0.005)& (0.016)& (0.003)& (0.011) \\
  	\hline
  \end{tabular}
  \label{tab:ADS_binary}
\end{table}

\section{Discussion}
\label{sec:conc}
In this paper, we study tensor completion under non‐random missingness by modeling observation probabilities as a parametric function of the latent tensor. Our framework accommodates diverse data types, including continuous, binary, and counts.
For estimation, we develop an alternating maximization algorithm and derive non-asymptotic error bounds for the estimator at each iteration. We further propose a statistical procedure for testing whether missingness depends on tensor values, providing a formal assessment of missingness mechanism within our framework. 

There are several promising directions for future work. First, our framework can be extended to include auxiliary covariates, such as user or item attributes in recommender systems. These covariates could be incorporated into both the tensor decomposition model and the missingness mechanism model. 
Second, in our current framework, the observation probability function $g_{\btheta}$ is set to a parametric function. A natural extension is to consider a more flexible nonparametric model, with estimation carried out using methods such as kernel smoothing or splines. 
Finally, while our current inference targets the missingness parameter $\btheta$, an important extension is to develop inference procedures for the tensor entries themselves, possibly using techniques in \citet{xia2022inference,ma2024statistical}. 


\bibliographystyle{chicago}
\begingroup
\baselineskip=20pt
\bibliography{MNAR}
\endgroup
\end{document}